\documentstyle[12pt]{article}
\newcommand{\DerA}{\mbox{Der${\cal A}$}}

\newcommand{\Cstar}{\mbox{$C^*$}}

\newcommand{\calA}{\mbox{${\cal A}$}}
\newcommand{\kerG}{\mbox{${\rm ker}{\bf G}$}}

\newcommand{\za}{\mbox{${\cal Z}({\cal A})$}}

\newcommand{\EG}{\mbox{${{\cal E}_{\bf G}}$}}
\newcommand{\EGbar}{\mbox{${{\bar{\cal E}}_{\bf G}}$}}

\begin{document}

\title{Noncommutative unification of general relativity with
quantum mechanics and canonical gravity quantization}
\author{Michael Heller\thanks{Correspondence address: ul.
Powsta\'nc\'ow Warszawy 13/94, 33-110 Tarn\'ow, Poland. E-mail:
mheller@wsd.tarnow.pl} \\ Vatican Observatory, \\ V-1200 Vatican
City State
\and Wies{\l}aw Sasin \\
Warsaw University of Technology, Institute of Mathematics \\
Plac Politechniki 1, 00-661 Warsaw, Poland}

\maketitle

\begin{abstract}
The groupoid approach to noncommutative unification of general
relativity with quantum mechanics is compared with the canonical
gravity quantization.  It is shown that by restricting the
corresponding noncommutative algebra to its (commutative)
subalgebra, which determines the space-time slicing, an
algebraic counterpart of superspace (space of 3-metrics) can be
obtained. It turns out that when this space-time slicing emerges
the universe is already in its commutative regime. We explore
the consequences of this result.
\end{abstract}


\section{Introduction}
In recent years a new approach has appeared to the quantization
of gravity, the one based on noncommutative geometry. The idea
is to make space-time a noncommutative space (which is
essentially nonlocal) with the hope that in this way at least
some major obstacles to the gravity quantization could
eventually be overcome. There are many attempts in this
direction \cite{one}. In \cite{HSL} we have followed Connes
\cite[p. 99]{Connes} who, in order to make a space $X$
noncommutative defines a noncommutative algebra not directly on
$X$ but rather on a groupoid over $X$. This approach, which has
been further developed in the series of works \cite{ourmodel},
will be called a groupoid approach to the unification of general
relativity and quantum mechanics.
\par 
The aim of the present paper is to compare the groupoid approach
with the canonical gravity quantization \cite{four}, which can
be thought of as a ``reference point'' for other methods of
quantizing gravity. The groupoid approach is ``more radical'' in
the sense that in this approach the noncommutative counterpart
of the differential structure is quantized whereas in the
canonical method three-metrics play the role of ``quantization
variables''. We show that in spite of this difference the
superspace formulation of general relativity (which could be
regarded as a prerequisite of the canonical quantization) can be
obtained from the groupoid approach if the corresponding
noncommutative algebra is restricted to its commutative
subalgebra which determines a suitable slicing of space-time.
Consequently, in the groupoid approach when the space-time
slicing appears gravity is already in its ``classical
(non-quantum) regime''.  However, this conclusion could follow
from a simplification inherent in our model, and could
eventually be avoided if one considers a more general module of
the noncommutative counterpart of vector fields (the module of
derivations of a given algebra).
\par 
We organize our material in the following way. To make the paper
self-contained and to fix our notation, in Section 2, we give a
summary of the groupoid approach to noncommutative unification
of general relativity with quantum mechanics. In Section 3, we
define the noncommutative algebraic counterpart of the standard
concept of superspace (the space of three metrics). The
comparison of the canonical gravity quantization with the
groupoid approach is done in Section 4, and some conclusions and
comments are collected in Section 5.
\par

\section{Basic ideas of the model}
The main idea of the groupoid approach to the unification of
general relativity and quantum mechanics is to forget, in the
very beginning, the concept of space-time and start with the
abstract space $G = E \times \Gamma $, where $E$ is the total
space of a principal fibre bundle, and $\Gamma $ its structural
group such that the orbits of the action of $\Gamma $ on $E$
form a smooth manifold $M$ interpreted as space-time (this
construction can eventually be generalized to the category of
differential spaces of constant dimension, see \cite{HelSas1}).
We endow $G$ with the groupoid structure. In the present paper,
for the sake of concreteness, we shall assume that $E$ is the
total space of the frame bundle over a space-time manifold $M$,
and $\Gamma $ the group SO(3,1). Of course, $M = (G/\rm
{SO(3,1)})/\rm{SO(3,1)})$.  Then one defines the algebra as the
(intrinsic) direct sum
$$
\calA = \calA_{const} \oplus C^{\infty }_c(G, {\bf C})
$$
where $\calA_{const} = pr^*(C^{\infty }(M, {\bf C}))$, and
$C^{\infty }_c(G, {\bf C})$ is the family of smooth compactly
supported complex valued functions on $G$. The multiplication in
the algebra \calA \ is defined in the following way: (1) if $a,b
\in C^{\infty }_c(G, {\bf C})$, their multiplication is the
convolution $(a * b)(\gamma) =
\int_{G_p}a(\gamma_1)b(\gamma_2)$, where $\gamma = \gamma_1
\gamma_2$ with $\gamma , \, \gamma_1, \, \gamma_2 \in G_p$, $G_p$
being the fiber in $G$ over $p \in E$; integration is with
respect to the Haar measure; (2) if $a, b \in
\calA_{const}$ they are multiplied in the usual way, i. e., $a *
b = a \cdot b$; (3) if $a \in \calA_{const }$ and $b\in
C^{\infty }_c(G, {\bf C})$, one sets $(a * b)(\gamma ) = (b *
a)(\gamma ) = k(p)\int_{G_p}b(\gamma^{-1}_1 \gamma)$ where $k(p)
= \int_{G_p}a(\gamma_1)$.
\calA \ is evidently a noncommutative algebra. We also define
the involution of $a \in \calA $ by $a^*(\gamma ) =
\overline{a(\gamma^{-1})}$ where $\gamma = (p,g),\,p \in E,\, g
\in \Gamma $.\footnote{One should notice that we have corrected
the definition of the algebra \calA \ as compared with our
previous works (see \cite{HSL,ourmodel}). This corection does
not change our previous results.}
\par
Let us also define the subalgebra $\calA_{proj} =
\pi_M^*C^{\infty }(M, {\bf C}) \subset \calA_{const}$. It plays
the important role in our model since by restricting the algebra
\calA \ to the subalgebra $\calA_{proj }$ we recover the
space-time manifold of general relativity.
\par
Let us consider the set \DerA \ of all derivations of the
algebra \calA . \DerA \ is a \za -module, where \za \ denotes
the center of \calA , and can be regarded as a noncommutative
counterpart of vector fields. In the following, we shall
consider a noncommutative differential geometry as defined by
the \za - submodule $V$ of \DerA \ such that $V = V_E \oplus
V_{\Gamma }$ where $V_E$ and $V_{\Gamma }$ are derivations of
\calA \ parallel to  $E$ and $\Gamma $, respectively (this is
only a simplifying assumption which in the general case should
be relaxed).
\par
First, we define a {\em metric\/} on the ${\cal Z}({\cal A}
)$-submodule $V$  as a \za -bilinear non-degenerate symmetric
mapping $g :V\times V\rightarrow\calA$, and for our model we
choose the following metric adapted to the product structure of
$V$
\begin{equation}
g=pr_E^{*}g_E+pr_{\Gamma}^{*}g_{\Gamma} \label{metric}
\end{equation}
where $g_E$ and $g_{\Gamma}$ are metrics on $ E$ and $\Gamma$,
respectively, and $pr_E$ and $pr_{\Gamma }$ are the obvious
projections. It turns out that the ``vertical component''
$pr^*_{\Gamma }g_{\Gamma }$ of the metric $g$ is essentially
unique (this is true for a broad class of derivation based
noncommutative differential calculi, see \cite{MadMou}), whereas
the ``parallel component'' $pr^*_Eg_E$ of $g$ is a lifting of
the Lorentz metric in space-time $M$ (see also \cite{Towards}).
\par
Now, with the help of the Koszul formula, we define the linear
connection; then the curvature and the usual Ricci operator
${\bf R}:V\rightarrow V$ which is the counterpart of the Ricci
tensor with one index up and one index down (for details see
\cite{HSL}). In this way, we have all quantities needed to write
the {\em noncommutative Einstein equation\/}
\begin{equation}
{\bf G} = 0\label{R2}
\end{equation}
where ${\bf G} = {\bf R}+2\Lambda {\bf I}$ with ${\bf R}$ being
the Ricci operator, $\Lambda$  a constant related to the usual
cosmological constant, and ${\bf I}$ the identity operator.
Because of the form of metric (\ref{metric}) ${\bf G}$ also
assumes the form ${\bf G}_E + {\bf G}_{\Gamma }$ (with obvious
meaning of symbols).
\par
The set ${\rm ker}{\bf G} = {\rm ker}{\bf G}_E \oplus {\rm
ker}{\bf G}_{\Gamma }$ is a $\za $-submodule of $V$ and
represents a solution of eq. (\ref{R2}). Because of the
uniqueness of the metric $pr^*_{\Gamma }g_{\Gamma }$ the
equation ${\bf G}_{\Gamma} = 0$ should be solved for derivations
$v\in {\rm ker}{\bf G}_{\Gamma } \subset V_{\Gamma }$. The
equation ${\bf G}_E = 0$, as a ``lifting'' of the usual
Einstein's equation should be solved for the metric. All
derivations $v\in V_E$ satisfy it automatically (and all
derivations $v \in V_{\Gamma }$ satisfy it trivially, see
\cite{Towards}). 
\par
Let us consider the representation of the algebra ${\cal A}$ in
the Hilbert space ${\cal H}=L^2(G_q)$, $\pi_q:\calA\rightarrow
{\cal B}({\cal H})$, where ${\cal B}({\cal H})$ denotes an
algebra of bounded operators on ${\cal H}$ and $G_q$ is the
fiber of $G$ over $q
\in E$, given by the formula
\begin{equation}
(\pi_q(a)\psi )(\gamma )=\int_{G_q}a(\gamma_1)\psi
(\gamma_1^{-1}\gamma ),\label{Conrepr}
\end{equation}
with $\gamma =\gamma_1\circ\gamma_2,\;\gamma ,\gamma_
1,\gamma_2\in G_q, \,q\in E\,;\psi\in {\cal H},\;a\in {\cal A}$.
The integral is taken with respect to the Haar measure. The
completion of \calA\ with respect to the norm $$
\parallel a\parallel\,={\rm s}{\rm u}{\rm p}_{
q\in E} \parallel\pi_q(a)\parallel $$ is a $C^{*}$-algebra  (see
\cite[p. 102]{Connes}). We shall denote this algebra by ${\cal
E}$.
\par
We assume (as a separate axiom) that the dynamics of a quantum
gravitational system is described by the following equation
\begin{equation}
i\hbar\pi_q(v(a))=[F_v, \pi_q(a)] \label{dyneq}
\end{equation} 
for every $q\in\ E$, where $v\in {\rm k}{\rm e}{\rm r}{\bf G}$,
and $(F_v)_{v\in {\rm ker}{\bf G}}$ is a one-parameter family of
operators $F_v\in {\rm End}{\cal H}$ with ${\cal H}=L^2(G_q)$
such that $$ F_{\lambda_1v_1+\lambda_2v_2}=\lambda_1F_v+\lambda_
2F_2$$ for $v_1,\,v_2\in {\rm ker}{\bf G}
,\,\lambda_1,\,\lambda_ 2\in {\bf C}$.  We shall also assume
that $[F_v, \pi_q(a)]$ is a bounded operator.
\par
The fact that $v\in\kerG$ makes of eqs.  (\ref{R2}) and
(\ref{dyneq}) a ``noncommutative dynamical system''. We could
also say that noncommutative Einstein equation (\ref{R2}) plays
the role of a ``boundary condition'' for quantum dynamical
equation (\ref{dyneq}). To solve this system means to find the
set
\[\EG=\{a\in {\cal E}:i\hbar \pi_q(v(a))=[F_v, \pi_
q(a)],\forall v\in {\rm k}{\rm e}{\rm r}{\bf G}
\}.\]
It can be easily verified that it is a subalgebra of $ {\cal
E}$.

Let \EGbar\ be the smallest closed involutive subalgebra of the
algebra ${\cal E}$ containing \EG.  \EGbar\ is said to be {\em
generated by }\EG.  Since $ {\cal E}$ is a
\Cstar-algebra and every closed involutive subalgebra of a 
\Cstar-algebra is a \Cstar-algebra (see \cite[Sec.
1.3.3]{Dix}), \EGbar\ is also a \Cstar-algebra; it will be
called {\em Einstein \Cstar-algebra\/} or simply {\em Einstein
algebra}, and the pair $ (\EGbar,\kerG)$ -- {\em Einstein
differential algebra}.

Now, the idea is to perform quantization with the help the usual
C$^{*}$-algebraic method (see, for instance, \cite{Thirring},
\cite[chapter 9]{Emch}) with the Einstein
algebra \EGbar\ as our basic \Cstar-algebra. According to this
method, a quantum gravitational system is represented by \EGbar,
and its observables by Hermitian elements of \EGbar. If $ a$ is
a Hermitian element of $\EGbar$, and $\phi$ a state on
\EGbar \ then $\phi (a)$ is the expectation value of the
observable $ a$ when the system is in the state $\phi$.

It can be shown that this gravity quantization scheme correctly
reproduces the usual general relativity (on space-time) and
quantum mechanics (in the Heisenberg picture) when the algebra
${\cal A}$ is restricted to its center ${\cal Z} ({\cal A})$ (or
to some subset of ${\cal Z}({\cal A})$) (see
\cite{HSL,Towards}).

\section{Algebraic version of superspace}
First, let us recall the well known construction. Let Riem$(S)$
denote the space of all Riemannian metrics on a 3-manifold $S$,
and let Diff$(S)$ be the group of all orientation preserving
diffeomorphisms of $S$. For simplicity, we assume that $S$ is
closed (e. g., compact and without boundary).  We have the
action of Diff$ (S)$ on Riem$(S)$
\[{\rm D}{\rm i}{\rm f}{\rm f}(S)\times {\rm R}
{\rm i}{\rm e}{\rm m}(S)\rightarrow {\rm R}{\rm i} {\rm e}{\rm
m}(S)\] given by
\[(f,h)\mapsto f^{*}h.\]
The quotient space ${\cal S}(S)=\frac {{\rm R} {\rm i}{\rm
e}{\rm m}(S)}{{\rm D}{\rm i}{\rm f} {\rm f}(S)}$ is called {\em
superspace}.  Its global properties were studied by Fischer
\cite{Fischer} (see also \cite{Giulini}).
\par
In a particular coordinate system any metric $h\in$ Riem$(S)$
can be represented as a covariant metric tensor $h_{ij}(x)$ or
as a contravariant metric tensor $ h^{ij}(x),\,x\in S$. Then, as
shown by DeWitt \cite{DeWitt67}, there exists a metric on ${\cal
S}(S)$, called the {\em Wheeler-DeWitt metric}, which assumes
the form
\begin{equation}
G_{ijkl}=\frac 12h^{-1/2}(h_{
ik}k_{jl}+h_{il}h_{jk}-h_{ij}h_{kl}).
\label{eq1}\end{equation}
It has the signature $(-+++++)$ for each point of the
3-geometry.
\par 
Let us now consider a slicing
$(S_t)_{t\in T}$ of $M$ such that $S_t$ is diffeomorphic to $S$
for each $t \in T$. Let further $\calA_S \subset \calA $ be the
subalgebra of functions which are constant on
$pr^{-1}(S_t)_{t\in T}$, where $pr = pr_M \circ pr_E$ with
$pr_M: E \rightarrow M$ being the canonical projection, and let
us denote by $V_S$ the set of all derivations of \calA \ which
are invariant with respect to $\calA_S$, i. e., such that
$V_S(\calA_S) \subseteq \calA_S$. Evidently, we have $V_S
\subset V_E$. Let us notice that the subalgebra $\calA_{proj }$
can be equivalently defined in another way; namely as consisting
of functions of \calA \ which are constant on the
equivalence classes of fibres $G_p = pr^{-1}(x), \, pr_M(p)=x\in
M$. Two fibres $G_p$ and $G_q$, $p,q\in E$ are equivalent if
there is $g\in \Gamma $ such that $q=pg$. Now, it can be easily
seen that $\calA_S \subset \calA_{proj} \subset \za $. Indeed,
$pr^{-1}(x) \subset pr^{-1}(S)$ for every $x \in S$.
Consequently, the differential algebra $(\calA_S, V_S)$ is
commutative.  We denote the set of all metrics in the module
$V_S$ by Riem$(\calA_S)$. As an analogue of Diff$(S)$ we should
take the set Iso$ {\cal A}_S$ of all isomorphisms of $\calA_S$
into itself.  We have the action
\[{\rm I}{\rm s}{\rm o}(\calA_S)\times {\rm R}
{\rm i}{\rm e}{\rm m}(\calA_S)\rightarrow {\rm R}{\rm i}{\rm
e}{\rm m}(\calA_S)\] defined by
\[(f,h)\mapsto f^{*}h.\]
Any isomorphism $f:{\cal A}_S\rightarrow {\cal A}_ S$ induces
the mapping (which is also an isomorphism)
\[f^{\#}:V_S\rightarrow V_S\]
by
\[f^{\#}(v)(\alpha )=v(f^{*}\alpha )=v(\alpha
\circ f)\]
where $v\in$ $V_S$, $\alpha\in \calA $. Therefore, one has
\[(f^{*}h)(v_1,v_2)=h(f^{\#}v_1,f^{\#}v_2),\]
$v_1,v_2\in$ $V_S$, $h \in {\rm Riem}\calA_S$, and we can define
the superspace associated with the algebra  \calA \ as
\[{\cal S}(\calA):=\frac {{\rm R}{\rm i}{\rm e}
{\rm m}(\calA_S)}{{\rm I}{\rm s}{\rm o}({\cal A}_ S)}.\]
\par
We have the following conclusion:  By restricting the algebra
\calA \ to its subalgebra $\calA_S$ and
considering the set Riem$ ({\cal A}_S$) of all Riemannian
metrics in the \za -submodule $V_S$ one obtains the algebraic
counterpart of the standard concept of superspace.
\par

\section{Noncommutative gravity and canonical quantization}
We now briefly recollect the canonical method of quantizing
gravity to compare it with our approach.  Any space-time metric
can be locally written in the form
\begin{equation}\label{eq2}ds^2=-(N^2-N_iN^i)
dt^2+2N_idtdx^i+h_{ij}dx^idx^j,\end{equation} where $h_{ij}$,
$i,j=1,2,3$ is the metric tensor on the spacelike hypersurface
$S=$const, $N$ is called {\em lapse function\/}; it measures the
proper time separation between hypersurfaces $t=$const.  The
so-called {\em shift vector} $ N_i$ measures the deviation of
curves $x^i=$const from the normal to $ S$ (in the following we
use units such that $c=\hbar =1$).  The extrinsic curvature of $
S$ can be written as
\[K_{ij}=\frac 1{2N}[-\frac {\partial h_{ij}}{
\partial t}+2N_{i|j}],\]
where the stroke $``$$|$'' denotes covariant differentiation
with respect to the 3-metric $h_{ij}$. The momentum canonically
conjugated to $ h_{ij}$ is given by
\[\pi^{ij}=-h^{1/2}(K^{ij}-h^{ij}K),\]
where $K=K^i_i$.
\par
The classical Hamiltonian is
\begin{equation}\label{eq3}H=\int (NH_0+N_iH^
i)d^3x,\end{equation} where
\[H_0=G_{ijkl}\pi^{ij}\pi^{kl}-h^{1/2}(^3R-2\Lambda 
),\]
\[H^i=-2\pi^{ij}_{|j},\]
with $^3R$ being the scalar curvature of $h_{ ij}$ and $\Lambda$
the cosmological constant. By making the standard substitution:
$h_{ij}\mapsto h_{ ij},\,\pi^{ij}\mapsto -i\frac {\delta}{\delta
h_{ij}}$ $(\delta$ is the functional derivative) one obtains the
counterpart of the Schr\"odinger equation
\begin{equation}\label{eq4}\hat {H}\Psi =0.\end{equation}
The $\hat {H}_0$-part of this equation
\begin{equation}-[G_{ikl}\frac {\delta^2}{\delta 
h_{ij}\delta h_{kl}}+h^{1/2}(^3R-2\Lambda )]\Psi
[h_{ij}]=0.\label{eq5}\end{equation} is the celebrated {\em
Wheeler-DeWitt equation\/}. This is the fundamental equation for
the ``wave function of the universe'' $\Psi [h_{ij}]$ which is
the functional of the 3-metric (we do not take into account any
matter fields).
\par
We should emphasize that in the Wheeler-DeWitt approach it is
the 3-metric that is quantized (and the momentum canonically
conjugated to it), whereas in our approach the
``quantization variables'' are elements of the Einstein $
C^{*}$-algebra \EGbar . However, we can ask the question: what
would happen to the equations of our theory (eqs.  (\ref{R2})
and (\ref{dyneq})) if we restrict $\EGbar $ to $(\EGbar )_S$, i.
e.  if we go to the ``superspace limit''?
\par
Since $(\EGbar )_S \subset {\cal Z}(\EGbar )$ eq. (\ref{dyneq})
reduces to the trivial identity ($0 \equiv 0$) and hence it
becomes insignificant. We are left with eq. (\ref{R2}) which, in
this case, is reduced to the usual Einstein equations. In this
way, gravity decouples from quantum mechanics. This is an
important conclusion: if we go to the superspace limit quantum
gravity effects become negligible. In this process, the slicing
of space-time emerges, and consequently the concepts of time and
instantaneous spaces become meaningful. This means that we are
well beyond the Planck threshold in the non-quantum gravity
regime (see \cite{Emergence} where the emergence of time from
the noncommutative era has been studied).
\par
As it is well known, the Wheeler-DeWitt equation corresponds to
the stationary Schr\"odinger equation. Eq. (\ref{dyneq}) plays
the similar role in our approach since, for weak gravitational
fields it reduces to the Schr\"odinger equation (in the
Heisenber picture of quantum mechanics) \cite{Towards}. However,
one should not forget that the Wheeler-DeWitt equation is the
equation for three-metrics, whereas eq. (\ref{dyneq}) is the
equation for elements of the algebra \EGbar .

\section{Concluding remarks}
We have demonstrated that if in the groupoid approach to the
unification of general relativity and quantum mechanics,
proposed in \cite{HSL}, the algebra $\calA = \calA_{proj} \oplus
C^{\infty}_c(G, {\bf C})$ is
restricted to its subalgebra $\calA_S $, consisting of functions
constant on $pr^{-1}(S_t)_{t\in T}$, where $(S_t)_{t\in T}$ is a
time slicing of space-time $M$, one obtains the superspace
formulation of general relativity.
\par
The important point is that our approach shows that at the level
where time slicing of space-time appears, quantum gravity
effects are already insignificant (i. e., gravity is too weak to
exhibit quantum effects, see above Section 4). This seems
reasonable since in the quantum gravity regime we would expect
some kind of ``foamy mixture'' of space and time which is
excluded by the well defined time slicing of space-time.
This conclusion could be the consequence of a
simplifying assumption incorporated into our model, namely that
our noncommutative differential algebra is based on the \za
-submodule $V$ of Der\calA \ such that $V = V_E \oplus V_{\Gamma
}$ where $V_E$ and $V_{\Gamma }$ are submodules of derivations
parallel to $E$ and $\Gamma $, respectively. In this model
``geometry parallel to $E$'' is, in principle, responsible for
gravity effects and ``geometry parallel to $\Gamma $'' is
responsible for  quantum effects. The fact that we have
neglected ``mixed terms'' (those coming both from $V_E$ and
$V_{\Gamma }$) means that in our model gravity is ``weakly
coupled'' to quantum effects.  Consequently, if we restrict the
algebra $\calA  $ to its subalgebra $\calA_S $ (this
restricting essentially means that slicing of space-time enters
the scene) all terms parallel to $\Gamma $ automatically are
switched off. Such terms would be responsible for a
``fluctuating slicing'' of space-time which could be enough for
an approximate validity of the canonical quantization of
gravity. The decisive step in checking this hypothesis would be
to construct a counterpart of our model based on a more general
module of derivations. 
\par
The analogous situation occurs in the canonical quantization
approach. One begins with the sliced classical space-time (with
no quantum effects). Then one performs the canonical
quantization, as the result of which 3-geometries begin to
fluctuate, and the sliced regime of space-time becomes
``fuzzy''.
\par
As it is well known, when Einstein's equations are formulated as
a constrained Hamiltonian system, the Hamiltonian constraint and
the equations of motion determine the evolution of three-metrics
in superspace, and the momentum constraint implies that the
Hamiltonian flow is orthogonal (in the Wheeler-DeWitt metric) to
the orbits of the diffeomorphism group (although these two
directions need not be disjoint \cite{Giulini}). Since in our
algebraic approach the submodule $V_S$ corresponds to the family
of vector fields on the superspace ${\cal S}(\calA)$ the above
mentioned regularities should be reflected in the structure of
this submodule.
\par
\vspace{1cm}
\noindent
ACKNOWLEDGMENT
\par
The authors express their gratitude to the late Professor
Jacques Demaret and Dr. Marek Biesiada for many discussions
which have significantly contributed to improving this paper.
\par

\end{document}